\documentclass[preprint* published]{JHEP3} 

\usepackage{epsfig,multicol}

\newcommand\fverb{\setbox\pippobox=\hbox\bgroup\verb}
\newcommand\fverbdo{\egroup\medskip\noindent%
			\fbox{\unhbox\pippobox}\ }
\newcommand\fverbit{\egroup\item[\fbox{\unhbox\pippobox}]}
\newbox\pippobox

\usepackage{amsbsy,amssymb,latexsym}

\newcommand{\nn}{\nonumber}
\newcommand{\e}{{\rm e}}

\newcommand{\del}{\delta}

\newcommand{\al}{\alpha}

\newcommand{\1}{\mathbb I}

\title{M-theory on a Time-dependent Plane-wave}

\author{Makoto Sakaguchi and Kentaroh Yoshida\\
Theory Division, High Energy Accelerator Research 
Organization (KEK),\\
Tsukuba, Ibaraki 305-0801, Japan.\\
	E-mail: \email{Makoto.Sakaguchi@kek.jp}, \email{kyoshida@post.kek.jp}}
\received{February 20, 2001} 		
\revised{May 1, 2001}
\accepted{November 27, 2001}		

\preprint{\hepth{0309025}}	

\abstract{We propose a matrix model on a homogeneous
plane-wave background with 20 supersymmetries.  This background is
anti-Mach type and is equivalent to the time-dependent background. We
study supersymmetries in this theory and calculate the superalgebra.
The vacuum energy of the abelian part is also calculated.  In addition
we find classical solutions such as graviton solution, fuzzy sphere and 
hyperboloid.}


\begin{document} 


\section{Introduction}

Superstring theories and M-theory can be consistently realized on
various classical solutions of supergravity theories.  However, it is
difficult to concretely study theories on these backgrounds because of
complicated interaction terms.  In the recent progress of string theory,
plane-wave backgrounds are focused upon. The type IIB pp-wave solution
was found \cite{BFHP1} and it was shown that this background can be
described as the approximation of the $AdS$ space geometry around a
certain null geodesics \cite{BFHP2,BFHP3} via the Penrose limit
\cite{P}.  The superstring theory on this background \cite{M,MT} is
exactly solvable in the Green-Schwarz formulation, in spite of the
presence of the Ramond-Ramond fluxes.  Moreover, this string theory was
utilized for the study of the $AdS$/CFT correspondence at the stringy
level \cite{BMN} beyond the supergravity analysis.

On the other hand, the matrix model on the eleven-dimensional pp-wave
background was proposed by Berenstein, Maldacena and Nastase \cite{BMN},
which is closely related to a supermembrane theory on this background
\cite{DSR,SY1} via the matrix regularization \cite{dWHN}.  This pp-wave
matrix model includes mass terms and the Myers term \cite{Myers}, and
hence one can expect interesting physics intrinsic to the pp-wave
case. Because of mass terms, all of the flat directions are completely
lifted up. So it might be expected that a single supermembrane should be
stabilized. In addition, the presence of the Myers term leads to many
classical solutions living only on the pp-wave. These solutions are
fuzzy sphere type ones. In particular, the fuzzy sphere solution with
zero energy exists, and it can appear in the classical vacuum of the
system. As the result, the vacuum of this matrix model is very enriched.

Thus, superstring theories and M-theory on plane-wave backgrounds are
quite interesting and so have been intensively studied.  However, most
of the studies have considered the time-{\it independent} backgrounds.
A recent interest in supergravity, superstring theories and M-theory is
to include the time-{\it dependent} background in our consideration.  In
general, it is much difficult to treat the time-dependent backgrounds
and so one can hardly study string theories and M-theory on such
backgrounds. Under such a circumstance, the time-dependent plane-wave
background was considered by Papadopoulos, Russo and Tseytlin
\cite{PRT}, and they showed that the string theory on this background is
exactly solvable.  The solvability of this model comes from
supersymmetries which are preserved as a characteristic of plane-wave
type backgrounds.  Thus, we can find that the plane-wave backgrounds are
very available to study the physics on time-dependent backgrounds.

The main focus of this paper is to study a matrix model on a
homogeneous plane-wave background and study various features such as
supersymmetries, vacuum structure and classical solutions in this
model. The homogeneous plane-wave background \cite{BO} we will consider
here leads to the time-dependent background through a coordinate
transformation. Hence the study of the matrix model on this homogeneous
background is equivalent to that on the time-dependent background.  On
the other hand, this background can be obtained from an M-theory
G$\ddot{\rm o}$del universe\footnote{The G$\ddot{\rm o}$del universe
was originally proposed in the paper \cite{Godel}} via the Penrose limit
and this is also an anti-Mach type solution\footnote{M-theory
G$\ddot{\rm o}$del universe can be constructed by lifting up a maximally
supersymmetric G$\ddot{\rm o}$del solution in the minimal supergravity
in five dimensions \cite{Gaunt}.  Also, string theories on G$\ddot{\rm
o}$del universe and homogeneous plane-wave background are discussed in
\cite{HT} and \cite{BOPT}, respectively.} \cite{aM}.  Therefore it can
be expected that the study here should be closely connected to the
M-theory on the G\"odel universe.

Our paper is organized as follows: In Section 2, we will briefly review
a homogeneous plane-wave background which leads to a time-dependent
background. In Section 3, the action of the matrix model on this
background will be proposed.  Section 4 is devoted to the study of the
vacuum structure of the abelian part of the matrix model.  In Section 5,
examples of classical solutions in this model will be presented. We will
discuss graviton solutions in our model and find fuzzy sphere and
hyperbolic type solutions. Section 6 is devoted to a conclusion and
discussions.  In Appendix A, we will discuss a matrix model on the
general homogeneous plane-wave. Appendix B is devoted to the detail
study of the energy spectrum.

\section{Time-dependent Homogeneous Plane-wave} 

In this section we will briefly review homogeneous plane-wave
backgrounds, which are Cahen-Wallach space \cite{CW} with rotation terms. 
This type of backgrounds are related to time-dependent homogeneous 
plane-waves via time-dependent coordinate transformations. 
These are solutions of the eleven-dimensional supergravity and 
the family of them is given (in the Brinkmann coordinates) by
\begin{eqnarray}
\label{2.1}
ds^2 &=& -2dx^+dx^- + \mu^2A_{ij}x^ix^j(dx^+)^2 + 2\mu f_{ij}x^i dx^j dx^+
+ (dx^i)^2\,, \\
 F_4 &=& \frac{\mu}{3!} dx^{+}\wedge \xi_{ijk}\,dx^{ijk}\,, 
\label{2.2}
\end{eqnarray}
where $A_{ij}$ and $f_{ij}$ 
are constant symmetric and anti-symmetric 
matrices respectively, and $\xi_{ijk}$ is a completely anti-symmetric 
third-rank tensor.
The nonzero-components of spin connection for this background are 
\begin{eqnarray}
\omega^{+i} = A_{ij}x^j dx^+ + f_{ij}dx^{j}\,, \quad 
\omega^{ij} = - f_{ij}dx^+\,. 
\end{eqnarray}
The metric (\ref{2.1}) explicitly describes the
time-dependent background if $A_{ij}$ and $f_{ij}$ do not commute, as
mentioned in \cite{BO}. In particular, 
the above type of plane-wave backgrounds is related to the explicitly
time-dependent metric
\begin{eqnarray}
\label{time}
ds^2 = -2dx^+dx^- + \left(\e^{x^+f}A_0\e^{-x^+f}\right)_{ij}z^iz^j
(dx^+)^2 + (dz^i)^2\,. 
\end{eqnarray} 
By the use of the coordinate transformation of transverse directions 
\begin{eqnarray}
z^i \longrightarrow w^i=\left(\e^{-x^+ f}\right)_{ik}z^k\,,  
\end{eqnarray}
the metric (\ref{time}) takes the stationary form
\begin{eqnarray}
ds^2 = -2dx^+dx^- + \left((A_0)_{ij} - f_{ik}f_{kj}\right)w^iw^j
(dx^+)^2 + (dw^i)^2 - 2w^if_{ik}dw^kdx^+\,,
\end{eqnarray}
and we see that the {\it time-dependent} plane-wave (\ref{time}) can
be mapped to the {\it stationary} plane-wave as in (\ref{2.1}). 
  
In this paper, we concentrate on the special case which admits 20
supersymmetries \cite{BMO}, where the matrices $A_{ij}$ and $f_{ij}$ are
given by
\begin{eqnarray}
\label{2.7}
 A_{22} = 4(P^2 -1)\,, \quad A_{33} = A_{44} = -1\,, \quad 
f_{12} = -P\,,
\end{eqnarray} 
and otherwise zero. The $\xi_{ijk}$ is written as 
\begin{eqnarray}
\label{2.8}
\xi_{129} = -2\,, \quad \xi_{349} = 2P\,, \quad \xi_{256} =
 -2(1-P^2)^{1/2}\,, \quad \xi_{278} = -2(1-P^2)^{1/2}\,.
\end{eqnarray}
Notably, the above $A_{ij}$ and $f_{ij}$ do not commute, and so this
background is equivalent to the time-dependent background.  We can
obtain this background from the M-theory G$\ddot{\rm o}$del universe by
the use of the Penrose limit \cite{P}. The M-theory G$\ddot{\rm o}$del universe
is constructed by lifting up a maximally supersymmetric G$\ddot{\rm
o}$del solution of minimal five-dimensional supergravity \cite{Gaunt} 
($\beta$ is an arbitrary parameter): 
\begin{eqnarray}
&& ds^2_{5{\rm G}} = - (dt + \beta(r_1^2d\phi_1^2 + r_2^2d\phi_2^2))^2 +
 dr_1^2 
+ r_1^2d\phi_1^2 + dr_2^2 + r_2^2d\phi_2^2\,, 
\end{eqnarray}
as the product space with the six-dimensional flat space $\mathbb{R}^6$: 
\begin{eqnarray}
ds^2_{\rm M} = ds^2_{5\rm{G}} + dz^2 + \sum_{i=5}^9(dx^i)^2\,,
\end{eqnarray}
where this eleven-dimensional background is supported by 
the four-form field-strength:
\begin{eqnarray}
F_{r_1\phi_1 56} = F_{r_1\phi_1 78} = F_{r_1\phi_1 9z} = 
F_{r_2\phi_2 56} = F_{r_2\phi_2 78} = F_{r_2\phi_2 9z} = -2 \beta\,. 
\end{eqnarray}
The plane-wave solution (2.1) and (2.2) admits 20 supersymmetries (not
maximal 32), which are inherited from that of the M-theory G\"{o}del
background (2.7), and its Killing spinor was explicitly presented in
\cite{BMO}. That is, this background admits extra Killing spinors
satisfying
\begin{eqnarray}
&& [
\gamma_i\theta^2
+6\theta\gamma_i\theta
+9\theta^2\gamma_i
+6\gamma_i[\theta,\phi]
+18[\theta,\phi]\gamma_i 
-144(A_{ij}\gamma^j+f_{ij}f_{jk}\gamma^k)
]\epsilon=0\,,\nonumber\\&&
\epsilon(x^+)=e^{\frac{x^+}{2}(\phi+\frac{1}{6}\theta)}\epsilon_0
\end{eqnarray}
where we have defined as 
\begin{eqnarray}
\theta \equiv \frac{1}{3!}\xi_{ijk}\gamma^{ijk}\,,\quad 
\phi \equiv \frac{1}{2}f_{ij}\gamma^{ij}\,,
\end{eqnarray} 
and $\epsilon_0$ is a constant spinor.
It is worth noting that the parameter $P$ characterizes the null
geodesic in taking the Penrose limit.  The null geodesic is
chosen to have a momentum $P$ along the six directions transverse to the
five-dimensional G$\ddot{\rm o}$del metric. If we consider the $P=0$
case, then the null geodesic is in the G$\ddot{\rm o}$del space in five
dimensions.

\section{Action of Matrix Model on a Time-dependent Plane-wave} 

Here we shall discuss the matrix model on the general homogeneous
plane-wave background (anti-Mach type space) whose metric is (\ref{2.1})
with (\ref{2.7}) and the four-form flux (\ref{2.2}) with (\ref{2.8}) is
equipped.  By generalizing the result of work \cite{Pope}, we can 
propose the action of matrix model in the light-cone gauge as follows:
\begin{eqnarray}
S &=& \int\!d\tau\,\frac{1}{2}{\rm Tr}\Biggl[(D_{\tau}X^i)^2 +
 \frac{1}{2}[X^i,X^j]^2 
+ i\psi^TD_{\tau}\psi -
 \psi^{T}\gamma_i[X^i,\psi]  \nn \\
&& \qquad -4\mu^2(1-P^2)(X^2)^2 - \mu^2(X^3)^2 -\mu^2(X^4)^2 \nn \\ 
&& \qquad + 2\mu P X^1D_{\tau}X^2 - 2\mu P X^2D_{\tau}X^1 
+ \frac{\mu}{2}i\psi^T \left(
W - P\gamma^{12}
\right) \psi \nn \\
&& \qquad - \frac{4}{3}i\mu \cdot 3!\bigl\{
X^{[1}X^2X^{9]} 
- P X^{[3}X^4X^{9]} \nn \\
&& \qquad + (1-P^2)^{1/2} X^{[2}X^5X^{6]}  
+ (1-P^2)^{1/2} X^{[2}X^7X^{8]}
\bigr\}\Biggr]\,, \\
W &=& \gamma^{129} - P \gamma^{349} + (1-P^2)^{1/2}\gamma^{256} + 
(1-P^2)^{1/2}\gamma^{278}\,,
\end{eqnarray}
where $\psi$ is the 16-components $SO(9)$ spinor and 
$\gamma^i$'s are the $SO(9)$ gamma matrices. The $\tau$ is the
world-volume time and the covariant derivative $D_{\tau}$ is defined as $
 D_{\tau}\ast = \partial_{\tau}\ast -i[\omega,\ast]$\,.
The gauge connection $\omega$ describes the area preserving
diffeomorphism in terms of membrane theory\footnote{We restrict
ourselves to the closed membrane case. It is an interesting issue 
to study open supermembrane theories on a time-dependent background.}. 
It is an easy task to generalize the above action to the background
(\ref{2.1}) with (\ref{2.2}),
and present the matrix model on the general homogeneous
plane-wave (anti-Mach) background. This generalization will be 
discussed in the appendix \ref{general:app}. 

The above matrix model includes the Myers terms due to the presence of
the constant four-form flux. Hence the fuzzy sphere solutions should
appear as classical solutions as in the pp-wave matrix model
\cite{BMN}. These solutions will be discussed in detail later. 

Here we will discuss supersymmetries of the matrix model on the
time-dependent background. The 4 dynamical and 16 kinematical
supersymmetries are preserved in this background, and so 
the same number of supersymmetries should be preserved in the 
matrix model on this background. 

The transformation law of 4 dynamical supersymmetries is 
given by 
\begin{eqnarray}
\del_{\epsilon}X^i &=& i\psi^T\gamma^i\epsilon(\tau)\,, \quad 
\del_{\epsilon} \omega 
\;=\; i\psi^T\epsilon(\tau)\,, \quad \epsilon(\tau) \equiv 
\e^{\frac{\mu}{6}\left(W + 3P\gamma^{12}\right)\tau}P_1P_2\epsilon_0\,, \\ 
\del_{\epsilon}\psi &=& \Biggl[
D_{\tau}X^i\gamma_i -\frac{i}{2}[X^i,X^j]\gamma_{ij} 
+ 2\mu P (X^1\gamma^2 - X^2\gamma^1) 
\nn \\
&& - \frac{\mu}{6}X^i\gamma_i\left(W + 3P\gamma^{12}\right)
- \frac{\mu}{2}X^i\left(W - P\gamma^{12}\right)\gamma_i
\Biggl]\epsilon(\tau)\,,\nn
\end{eqnarray}
where $\epsilon_0$ is a constant spinor and 
we have introduced the projection operators: 
\begin{eqnarray}
P_1 &=& \frac{1}{2}\left(\1 - \gamma_{5678}\right)\,, \\ 
P_2 &=& \frac{1}{2}\left(\1 + c\gamma_{1956} - s\gamma_{1234}\right)\,,
\end{eqnarray}
where $c \equiv (1-P^2)^{1/2}$ and $s \equiv P$. 
The following properties: 
\begin{eqnarray}
P_1^2 = P_1\,, \quad P_2^2 = P_2\,, \quad P_1P_2 = P_2P_1\,.
\end{eqnarray}
are satisfied, and $P_1$ and $P_2$ are independent each other. 
Hence a quarter of dynamical supersymmetries (i.e., 4 dynamical supersymmetries) 
can survive. 

The transformation law of 16 kinematical supersymmetries is represented
by 
\begin{eqnarray}
\del_{\eta}X^i = \del_{\eta}\omega = 0\,, \quad 
\del_{\eta}\psi = \eta(\tau)\,, \quad 
\eta(\tau) = \e^{-\frac{\mu}{2}\left(W - P\gamma^{12}\right)\tau}\eta_0\,,
\end{eqnarray}
where $\eta_0$ is a constant spinor. 

We can calculate the superalgebra in our model by using the standard
Dirac bracket. In fact, our matrix model contains complicated fluxes and
the characteristic parameter $P$ is turned on, and hence its
superalgebra is expected to contain the generators of rotations (angular
momenta) and nontrivial terms intrinsic to our system. This interesting
topic will be studied in detail in another place \cite{SSY}.

\section{Vacuum Energy of the Abelian Part} 

In this section we will consider the abelian part of the matrix model
proposed in Section 3. This part can be exactly solved and so 
one can calculate the vacuum energy of this part. 
We will see below that this energy is negative. 

First, let us consider the $P=0$ case. 
The abelian part of the action is given by 
\begin{eqnarray}
S&=&\frac{1}{2}\int\!d\tau\,\Biggl[
(\dot{X}^i)^2 - 4\mu^2 (X^2)^2 - \mu^2 (X^3)^2 - \mu^2 (X^4)^2 
+ i\psi^{T}\dot{\psi} + \frac{\mu}{2}i\psi^{T}W_0\psi
\Biggr]\,.
\end{eqnarray}
where we introduced the notations 
$W_0 \equiv \gamma_{129} + \gamma_{256} + \gamma_{278}$ and 
$\dot{X}^i \equiv \partial_{\tau}X^i$. 

The dynamical supersymmetry transformation is given by 
\begin{eqnarray}
\del_{\epsilon}X^i &=& i\psi^T\gamma^i\epsilon(\tau)\,, 
\quad \epsilon(\tau) \equiv \e^{\frac{\mu}{6}W_0\tau}P_1P_2\epsilon_0\,, 
\nn \\
\del_{\epsilon}\psi &=& \left[\dot{X}^i\gamma_i - \frac{\mu}{6}
			 X^i\gamma_i W_0  
- \frac{\mu}{2}X^i W_0\gamma_i\right]
\epsilon(\tau)\,,
\end{eqnarray}
where $P_1$ and $P_2$ are written as 
\[
 P_1 = \frac{1}{2}\left(\1 - \gamma_{5678}\right)\,, \quad 
P_2 = \frac{1}{2}\left(\1 + \gamma_{1956}\right)\,.
\]
On the other hand, the kinematical supersymmetry transformation is 
described by 
\begin{eqnarray}
&&\del_{\eta}X^i = 0\,, \quad \del_{\eta}\psi = \eta(\tau)\,, 
\quad \eta(\tau) \equiv \e^{-\frac{\mu}{2}W_0\tau}\,.
\end{eqnarray}

Here we shall decompose the 16-components spinor $\psi$ as  
\begin{eqnarray}
\psi = \psi^{(++)} + \psi^{(+-)} + \psi^{(-+)} +\psi^{(--)}\,,
\end{eqnarray}
where each of the spinors $\psi^{(++)}$, $\psi^{(+-)}$, 
$\psi^{(-+)}$ and $\psi^{(--)}$ have only 4 non-trivial components
and satisfy the following two chirality
conditions: 
\begin{eqnarray}
\gamma_{5678}\psi^{(\pm \bullet)} = \pm \psi^{(\pm \bullet)}\,, \quad 
\gamma_{1956}\psi^{(\bullet \pm)} = \pm \psi^{(\bullet \pm)}\,.
\end{eqnarray}
Notably, the matrix $W_0$ commute with $\gamma_{5678}$ and
$\gamma_{1956}$, and hence the chirality is preserved under the action
of $W_0$ (for example, the chirality of $W_0\psi^{(++)}$ is the same with 
$\psi^{(++)}$). By using this decomposition, we can rewrite the 
fermionic action as 
\begin{eqnarray}
S_{\rm F} &=& \int\!d\tau\,\Biggl[\,
i\psi^{(++)T}\dot{\psi}^{(++)} + i\psi^{(+-)T}\dot{\psi}^{(+-)} \nn \\
&& + i\psi^{(-+)T}\dot{\psi}^{(-+)}  + i\psi^{(--)T}\dot{\psi}^{(--)} \nn \\ 
&& + \frac{\mu}{2}i\psi^{(++)T}\Pi\psi^{(++)} - 
\frac{\mu}{2}i\psi^{(+-)T}\Pi\psi^{(+-)} \nn \\
&& + \frac{3}{2}\mu i\psi^{(-+)T}\Pi\psi^{(-+)} 
 + \frac{\mu}{2}i\psi^{(--)T}\Pi\psi^{(--)}  
\Biggr]\,,
\end{eqnarray}
where the matrix $\Pi$ is defined as $\Pi \equiv \gamma_{256}$.

The vacuum energy of bosons comes from the massive directions:
$X^2$, $X^3$ and $X^4$, and the net contribution is expressed as 
\begin{eqnarray}
E_{\rm B} = \frac{\mu}{2}\left(2 + 1 +1\right) = + 2\mu\,.
\label{zero-b-P=0}
\end{eqnarray}
The vacuum energy of fermions are evaluated as (for example, by
following the procedure in the work \cite{NSY}) 
\begin{eqnarray}
E_{\rm F} = - \frac{\mu}{2}(1 + 1 + 1 + 3 ) = -3\mu\,.
\end{eqnarray}
Thus the total vacuum energy is given by 
\begin{eqnarray}
E_{\rm tot} = - \mu\,.
\end{eqnarray}
In conclusion, the vacuum energy of the abelian part 
is negative\footnote{The negative vacuum energy of the abelian part of 
the matrix model on less supersymmetric pp-waves is also shown in
\cite{KL}.}. 

Next, we will consider the $P\neq 0$ case. 
The action of the abelian part is written as 
\begin{eqnarray}
S &=& \frac{1}{2}\int\!d\tau\,\Biggl[
(\dot{X}^i)^2 - 4\mu^2(1-P^2)(X^2)^2 - \mu^2(X^3)^2 - \mu^2(X^4)^2 \nn \\
&& + 2\mu P (X^1\dot{X}^2 - X^2\dot{X}^1)  
+ i\psi^T\dot{\psi} + \frac{\mu}{2}i\psi^T\left(
W - P\gamma^{12}
\right)\psi
\Biggr]\,,
\end{eqnarray}
where the expression of $W$ in the $P\neq 0$ case is 
\begin{equation}
W = \gamma^{129} - s\gamma^{349} + c\gamma^{256} + c\gamma^{278}.
\end{equation}

The transformation law of 4 dynamical supersymmetries is given by 
\begin{eqnarray}
\del_{\epsilon}X^i &=& i\psi^T\gamma^i\epsilon(\tau)\,, \quad 
\epsilon(\tau) \equiv 
\e^{\frac{\mu}{6}\left(W + 3P\gamma^{12}\right)\tau}P_1P_2\epsilon_0\,, \\ 
\del_{\epsilon}\psi &=& \Biggl[
\dot{X}^i\gamma_i +2\mu P(X^1\gamma^2 - X^2\gamma^1)
\nn \\
& & - \frac{\mu}{6}X^i\gamma_i\left(W + 3P\gamma^{12}\right)
- \frac{\mu}{2}X^i\left(W - P\gamma^{12}\right)\gamma_i
\Biggr]\epsilon(\tau)\,,\nn
\end{eqnarray}
where $\epsilon_0$ is a constant spinor and the projection operators are
written as  
\begin{eqnarray}
P_1 &=& \frac{1}{2}\left(\1 - \gamma_{5678}\right)\,, \\ 
P_2 &=& \frac{1}{2}\left(\1 + c\gamma_{1956} - s\gamma_{1234}\right)\,.
\end{eqnarray}
The transformation law of 16 Kinematical supersymmetries is 
given in (3.7).

To begin with, we will consider the bosonic part. 
In the $P\neq 0$ case the analysis of 
$X^1$ and $X^2$ directions are nontrivial. We can find that 
the effect of $P$ is to shift the origin of oscillator $X^2$ and 
to rescale the continuous spectrum of $X^1$, 
as we will discuss in Appendix. Thus, 
the resulting zero-point energy for bosons 
is identical with that in the $P=0$ case:  
\begin{eqnarray}
E_{\rm B} = 2\mu\,.
\end{eqnarray}

Next let us consider the fermionic part and focus upon 
the fermion mass term: 
\begin{eqnarray}
\frac{\mu}{2}i\psi\left(W - P\gamma^{12}\right)\psi\,. 
\end{eqnarray}
We decompose the spinor $\psi$ according to the chirality in terms of 
the matrices $\gamma^{5678}$ and $c\gamma^{1956} + s\gamma^{1234}$
\footnote{The choice of chirality matrices is not relevant to the projection
operator $P_1$ and $P_2$. } as in
the $ P=0$ case. 
Remarkably, these matrices commute with $W-P\gamma^{12}$. We can easily
see this property if we rewrite the $W-P\gamma^{12}$ as 
\[
 W-P\gamma^{12} = \gamma^{129}\left(
\1 + (\1-\gamma^{5678})(c\gamma^{1956} + s\gamma^{1234})
\right)\,,
\]
where we have used the gamma matrices $\gamma^i$'s satisfy
$\gamma^{123456789} =\1$. Thus, the chiralities of spinor are preserved 
under the action of $W-P\gamma^{12}$ even in the $P\neq 0$ case. 
Now we can rewrite the fermion mass term as 
\begin{eqnarray}
\frac{\mu}{2}i\psi^{(++)T}\Pi'\psi^{(++)} 
+ \frac{\mu}{2}i\psi^{(+-)T}\Pi'\psi^{(+-)} 
+ \frac{3}{2}\mu i\psi^{(-+)T}\Pi'\psi^{(-+)}  
- \frac{\mu}{2}i\psi^{(--)T}\Pi'\psi^{(--)}\,, 
\end{eqnarray}
where $\Pi' \equiv \gamma^{129}$ and the spinors $\psi^{++}$,
$\psi^{+-}$, 
$\psi^{-+}$ and $\psi^{--}$ satisfy the following chirality conditions 
with respect to the matrices $\gamma^{5678}$ and $c\gamma^{1956} +
s\gamma^{1234}$: 
\begin{eqnarray}
\gamma^{5678}\psi^{\pm\bullet} = \pm \psi^{\pm\bullet}\,, \quad
(c\gamma^{1956} + s\gamma^{1234})\psi^{\bullet\pm} = \pm
\psi^{\bullet\pm}\,.
\end{eqnarray} 
Thus the vacuum energy of fermion part is represented by 
\begin{equation}
E_{\rm F} = - \frac{\mu}{2} \left(1 + 1 + 3 + 1\right) = - 3\mu\,.
\end{equation}
In conclusion, the net vacuum energy is 
\begin{equation}
E_{\rm tot} = -\mu\,.
\end{equation} 
Thus the zero-point energy is completely identical with that in the 
$P=0$ case and it is independent of the parameter $P$. 

We can derive the spectrum of the abelian part, and we might expect that
the resulting spectrum would agree with that in the linearized
supergravity in eleven dimensions around the time-dependent background
consider here (as is shown in case of the eleven-dimensional pp-wave
background \cite{KY}).

\section{Classical Solutions}

In this section we will study the classical solutions of our matrix
model. There are the extra terms with $f_{ij}$, and so we can expect
that graviton solutions would be modified.  Also, fuzzy sphere solutions
should exist due to the presence of the Myers term. 

In order to study classical solutions we need equations of motion.  We
will set $\omega = \psi$ = 0 for simplicity in the following
consideration.  By taking a variation of action, the equations of motion
are obtained as
\begin{eqnarray}
\ddot{X}^1 &=& -[[X^1,X^i],X^i] + 2\mu P\dot{X}^2 - 2i\mu(X^2X^9 - X^9X^2)\,,
 \\
\ddot{X}^2 &=& -[[X^2,X^i],X^i] - 2\mu P\dot{X}^1 - 4\mu^2(1-P^2)X^2 
-2i\mu(X^9X^1 - X^1X^9) \nn \\
&& -2i\mu(1-P^2)^{1/2}(X^5X^6 - X^6X^5 + X^7X^8 - X^8X^7)\,, \\
\ddot{X}^3 &=& -[[X^3,X^i],X^i] -\mu^2X^3 + 2i\mu P(X^4X^9 -
 X^9X^4)\,,\\
\ddot{X}^4 &=& -[[X^4,X^i],X^i] -\mu^2X^4 + 2i\mu P(X^9X^3 -
 X^3X^9)\,,\\
\ddot{X}^5 &=& -[[X^5,X^i],X^i] -2i\mu(1-P^2)^{1/2}(X^6X^2 - X^2X^6)\,,
 \\
\ddot{X}^6 &=& -[[X^6,X^i],X^i] -2i\mu(1-P^2)^{1/2}(X^2X^5 -
 X^5X^2)\,,\\
\ddot{X}^7 &=& -[[X^7,X^i],X^i] -2i\mu(1-P^2)^{1/2}(X^8X^2 -
 X^2X^8)\,,\\
\ddot{X}^8 &=& -[[X^8,X^i],X^i] -2i\mu(1-P^2)^{1/2}(X^2X^7 -
 X^7X^2)\,,\\
\ddot{X}^9 &=& -[[X^9,X^i],X^i] -2i\mu(X^1X^2 - X^2X^1) 
+ 2i\mu P(X^3X^4 - X^4X^3)\,.
\end{eqnarray}
We can evaluate the energy of the classical solution by
using the classical potential $V$ in our model:
\begin{eqnarray}
2V &=& {\rm Tr}\Bigl[-\frac{1}{2}[X^i,X^j]^2 + 4\mu^2(1-P^2)(X^2)^2 +
 \mu^2(X^3)^2 + \mu^2(X^4)^2 \nn \\ && \quad +
 \frac{4}{3}i\mu\cdot 3!\bigl(X^{[1}X^2X^{9]} -P X^{[3}X^4X^{9]} \nn \\ &&
 \quad + (1-P^2)^{1/2}X^{[2}X^5X^{6]} + (1-P^2)^{1/2}X^{[2}X^7X^{8]}
 \bigr) \Bigr]\,.
\end{eqnarray}

We will discuss below some types of classical solutions and evaluate the
classical energies by using the above classical equations of motion and
potential.

\subsection{Graviton Solutions} 

To begin with, we will consider the graviton solutions.  In the case of
a matrix model in flat space, the graviton solution is a free particle
moving with a constant velocity, while in the pp-wave case this solution
is modified to be described by an oscillator because of the presence of
mass terms in the action.  In our case the graviton solution is slightly
different from that in the usual pp-wave cases as we will explain.

Let us investigate the graviton solution by considering the diagonal
matrix. The graviton solutions for $X^5,~\cdots,~X^9$ are free particles 
as in flat space. For $X^3$ and $X^4$ the solution is a
couple of harmonic oscillators as in the usual pp-wave case because of
mass terms.  
The remaining parts are $X^1$ and $X^2$. These directions couple each
other as follows:
\begin{eqnarray}
\ddot{X}^1 = 2\mu P \dot{X}^2\,, \quad 
\ddot{X}^2 = -2\mu P \dot{X}^1 -4\mu^2(1-P^2)X^2\,.
\end{eqnarray} 
We can easily solve this system of equations and the 
solution is 
\begin{eqnarray}
X^1 &=& P x_2\sin(2\mu\tau) - P\frac{v_2}{2\mu}\cos(2\mu\tau) 
+ \left(1-P^2\right)c_0\tau + c_0'\,, \\
X^2 &=& x_2\cos(2\mu\tau) + \frac{v_2}{2\mu}\sin(2\mu\tau) 
- \frac{P c_0}{2\mu}\,.
\end{eqnarray}
The $x_2$ and $v_2$ are the initial position and velocity of $X^2$,
respectively.  The $c_0$ and $c_0'$ are arbitrary constants related to
the initial velocity and position of $X^1$, respectively.  It should be
remarked that the $X^1$ direction behaves as a harmonic oscillator in
spite of the absence of its mass term, because of the contribution of
the term $f_{ij}$. Notably, the $X^1$ behaves as a freely moving
particle in the $P=0$ case and as a harmonic oscillator in the $P=1$
case.  This result shows a specific feature of our model.

In addition, we can easily see that there exist a rotating solution:
\begin{eqnarray}
&& X^3(\tau) + iX^4(\tau) = (X^3(0) + i X^4(0))\e^{i\mu\tau}\,, 
\quad  [X^3(0),X^4(0)] = 0\,,
\end{eqnarray}
as in the matrix model on the pp-wave background \cite{BMN}. 
The energy of this solution is given by 
\begin{eqnarray}
2V = \mu^2 {\rm Tr}\left[
(X^3(0))^2 + (X^4(0))^2
\right]\,.
\end{eqnarray}

The above solutions describe the motion of a D-particle in our model.
For non-diagonal cases,
the D-particle can expand to form a higher dimensional object with two
dimensions (i.e., membrane) because of the Myers terms.  The expanded
fuzzy membrane solutions will be discussed in the next subsection.

\subsection{Fuzzy Sphere and Hyperboloid Solutions}

Here we would like to find fuzzy sphere type solutions by
imposing the ansatz:
\begin{eqnarray}
 X^1 = \al J^1\,, \quad X^2 = \beta J^2\,, \quad 
X^9 = \al J^3\,,
\label{FS}
\end{eqnarray}
where the $J^a~(a=1,2,3)$'s are the $SU(2)$ generators and satisfy the 
$SU(2)$ Lie algebra $[J^a,J^b]=i\epsilon_{abc}J^c$. By putting the
ansatz into the equations of motion, we obtain two conditions
\begin{eqnarray}
\al\beta^2 + \al^3 -2\mu\al\beta = 0\,, \quad 
\al^2\beta - \mu\al^2 + 2\mu^2(1-P^2)\beta = 0\,.
\end{eqnarray}
These algebraic equations can be easily solved, and we obtain two
nontrivial solutions: 
\begin{eqnarray}
&& \al^2 = -2\mu^2(1-P^2) + \frac{\mu^2}{2}(1 + \sqrt{1 +
 8(1-P^2)}\,)\,, \nn \\ 
&& \beta = \frac{3}{2}\mu - \frac{\mu}{2}\sqrt{1 + 8(1-P^2)}\,,
\label{fz1}
\end{eqnarray}
and 
\begin{eqnarray}
&& \al^2 = -2\mu^2(1-P^2) + \frac{\mu^2}{2}(1 - \sqrt{1 +
 8(1-P^2)}\,)\,, \nn \\
&& \beta = \frac{3}{2}\mu + \frac{\mu}{2}\sqrt{1 + 8(1-P^2)}\,,
\label{fz2}
\end{eqnarray}
and a trivial solution $\al=\beta=0$.  The first nontrivial
solution describes a fuzzy ellipsoidal sphere and the second one
represents a fuzzy hyperboloid\footnote{The ellipsoidal sphere and
hyperbolic type solutions were constructed in the pp-wave matrix model
by D.~Bak \cite{Bak}.} in the range $0\leq P \leq 1$\footnote{If we
consider the region $P<0$ for the first case, then the shape of the
solution becomes a fuzzy hyperboloid. On the other hand, if we consider
the region $P>1$ for the second case, then the solution becomes fuzzy
ellipsoidal sphere. }.  In the second case the value of $\al$ becomes
purely imaginary and the compact $SU(2)$ group is replaced with the
noncompact group $SO(2,1)$.  The energies of the fuzzy ellipsoidal
sphere (\ref{fz1}) and hyperboloid (\ref{fz2}) are represented by
\begin{eqnarray}
2V &=& 
\left(\al^2\beta^2 - \frac{4}{3}\mu\al^2\beta\right)
{\rm Tr}[(J^1)^2 + (J^3)^2] \nn \\
&& \quad 
+ \left(\al^4 + 4\mu^2(1-P^2)\beta^2 - \frac{4}{3}\mu\al^2\beta\right)
{\rm Tr}(J^2)^2
 \nn \\
&=& - \al^2\beta^2/N , 
\end{eqnarray}
where we have normalized the trace for the $N$-dimensional
representation of $SU(2)$ generators as
\begin{eqnarray}
{\rm Tr}(J^aJ^b)= \frac{1}{N}\del^{ab}\,.
\end{eqnarray}
The energy for the fuzzy ellipsoidal
sphere (\ref{fz1}) is negative while that for the fuzzy
hyperboloid (\ref{fz2}) is positive.

Other fuzzy solutions are given by 
\begin{eqnarray}
&& X^2 = c \cdot\beta J^1\,, \quad 
X^5 = c \cdot\al J^2\,, \nn \\
&& X^6 = c\cdot \al J^3\,, \quad 
\mbox{otherwise zero}\,, 
\label{FS1}
\end{eqnarray}
and 
\begin{eqnarray}
&& X^2 = c\cdot\beta  J^1\,, \quad 
X^7 = c\cdot \al J^2\,, \nn \\
&& X^8 = c\cdot \al  J^3\,, \quad 
\mbox{otherwise zero}\,.
\label{FS2} 
\end{eqnarray}
In the same way, these two cases lead to the same conditions: 
\begin{eqnarray}
2c^3\left(\al^2\beta + 2\mu^2\beta - \mu\al^2\right) = 0\,, \quad 
c^3\al\left(2\mu\beta - \beta^2 -\al^2\right) =0\,.
\end{eqnarray}
The solutions of the above algebraic equations are 
\begin{eqnarray}
\al^2 = - 3\mu^2\,, \quad \beta = 3\mu\,,
\label{FS12.solution}
\end{eqnarray}
and the trivial solution $\al = \beta = 0$.  Thus, both of the nontrivial
solutions (\ref{FS1}) and (\ref{FS2})
with (\ref{FS12.solution}) are fuzzy hyperboloids and have the same energy given by
\begin{eqnarray}
2V &=& c^4 \cdot 
\left(4\mu^2 \beta^2 + \al^4 - \frac{4}{3}\mu \al^2\beta\right)
{\rm Tr}(J^1)^2 \nn \\
&& \quad + \left(
\al^2\beta^2 - \frac{4}{3}\mu\al^2\beta
\right) {\rm Tr}[(J^2)^2 + (J^3)^2] \nn \\
&=& c^4 \cdot \left(
4\mu^2\beta^2 + \al^4 - 4\mu\al^2\beta + 2\al^2\beta^2
\right)/N\,, \nn \\
&=& 27 \mu^4 c^4 /N\,.
\end{eqnarray}

We also can apply the above consideration to the $X^3$, $X^4$ and $X^9$
directions by supposing the following ansatz:
\begin{eqnarray}
X^3 = \al J^1\,, \quad X^4 = \al J^2\,, \quad 
X^9 = \beta J^3\,, \quad \mbox{otherwise zero}\,.
\label{FS3}
\end{eqnarray} 
By inserting this ansatz into the equations of motion, we can obtain the
two constraint conditions:
\begin{eqnarray}
 \mu\al^2 P + \al^2\beta = 0\,,\quad \al^3 + \al\beta^2 + \mu^2\al 
+ 2\mu P\al\beta
 = 0\,.
\end{eqnarray}
These equations can be solved and the solution is 
\begin{eqnarray}
\label{fz4}
 \al^2 = - \mu^2(1 - P^2)\,,  \quad 
\beta = - \mu P\,.
\end{eqnarray}
This is also fuzzy hyperboloid solution, whose energy is
represented by 
\begin{eqnarray}
2V &=&  
\left(\al^2\beta^2 + \mu^2\al^2 + \frac{4}{3}\mu P\al^2\beta\right)
{\rm Tr}[(J^1)^2 + (J^2)^2] + \left(\al^4 + \frac{4}{3}\mu P\al^2\beta\right) {\rm Tr}(J^3)^2
\nn \\
&=& - \mu^4 (1-P^2)^2/N \; \leq 0\,.
\end{eqnarray}
As the result, the energy of the fuzzy hyperboloid solution has negative
energy. 

Now we will discuss the range of the parameter $P$. For example, if we
consider the $P^2 > 1$ case in the solution (\ref{fz4}), then the values
of $\al$ and $\beta$ give the ellipsoidal fuzzy sphere solution. If $P^2
< 1$, then the shape of the solution is hyperbolic.  This result implies
that the shape of the solution should be modified as the parameter $P$
changes.  It should be remarked that the parameter $P$ might run from 0
to 1. If P is out of this range, then the bosonic mass term would be
tachyonic and it seems that some trouble would appear.  However, such an
issue cannot be caused in the Rosen coordinates since it would be
intrinsic to the system described in terms of Brinkmann coordinates.

We will not discuss further classical solutions but it would be expected
that other interesting solution can be constructed in this model, and so
the construction of such solutions are an interesting future problem.
The stability of the fuzzy hyperboloids, which would be
non-supersymmetric, would be also interesting\footnote{The quantum
stability of fuzzy sphere is discussed in \cite{SY3}. }.

\subsection{BPS Equations} 

In the previous subsection, we have presented examples of classical
solutions.  Now we will consider the conditions which lead to
supersymmetric classical solutions in terms of BPS equations.

We can derive the BPS equations from the supersymmetric condition: 
$\del_{\epsilon}\psi =0$. 
The conditions preserving all of dynamical supersymmetries are given by 
\begin{eqnarray}
\label{BPS}
&& [X^2,X^9] = i\frac{4}{3}\mu X^1 \,, \quad 
 [X^9,X^1] = i\frac{4}{3}\mu X^2 \,, \quad 
 [X^1,X^2] = i\frac{4}{3}\mu X^9 \,,  \nn \\
&& [X^4,X^9] = -i s\cdot\frac{4}{3}\mu X^3\,, \quad 
[X^9,X^3] = -i s\cdot\frac{4}{3}\mu X^4\,, \quad 
[X^3,X^4] = -i s\cdot\frac{4}{3}\mu X^9\,, \nn \\
&& [X^5,X^6] = i c\cdot\frac{2}{3}\al\mu X^2\,, \quad 
[X^6,X^2] = i c\cdot\frac{4}{3}\mu X^5\,, \quad 
[X^2,X^5] = i c\cdot\frac{4}{3}\mu X^6 \,, \quad \nn \\
&& [X^7,X^8] = i c\cdot\frac{2}{3}\mu(1-\al) X^2\,, \quad 
[X^8,X^2] = i c\cdot\frac{4}{3}\mu X^7\,, \quad \quad 
[X^2,X^7] = ic\cdot\frac{4}{3}\mu X^8\,, \nn \\ 
&& D_{\tau}X^2 = - 2\mu s X^1\,, \quad 
D_{\tau}X^1 = 2\mu s X^2\,, \quad D_{\tau}X^4 = \mu X^3\,, \quad 
D_{\tau}X^3 = - \mu X^4\,, \nn \\
&& 
D_{\tau}X^5 =D_{\tau}X^6 = D_{\tau}X^7 = D_{\tau}X^8 = D_{\tau}X^9 = 0\,,
\end{eqnarray}
where a constant parameter $\al$ takes the value in the range $0< \al
<1$. The commutators which do not appear in the conditions (\ref{BPS})
should vanish (for example, $[X^1,X^5]=0$).  We note that the equations
(\ref{BPS}) are
not necessary
but sufficient conditions for the BPS conditions.
Because we are dealing with a non-maximally supersymmetric case,
some equations are satisfied trivially.

Now we can find a solution of the above BPS conditions (\ref{BPS}) 
represented by 
\begin{eqnarray}
&& X^1 = \left(\frac{4}{3}\mu J^2\right)\otimes \1\otimes \1 \otimes \1\,, \quad  
X^9 = \left(\frac{4}{3}\mu J^1\right)\otimes \left(-s\cdot\frac{4}{3}\mu J^1
\right) \otimes \1 \otimes \1\,, \nn 
\\
&& X^2 = \left(\frac{4}{3}\mu J^3\right) \otimes \1 \otimes 
\left(c\cdot\frac{4}{3}\mu
       J^3\right) \otimes \left(c\cdot\frac{4}{3}\mu J^3\right)\,, \nn \\
&& X^3 = \1\otimes \left(-s\cdot\frac{4}{3}\mu J^2\right) \otimes \1 \otimes \1\,, \quad X^4 = \1\otimes \left(-s\cdot \frac{4}{3}\mu J^3\right) \otimes \1 
\otimes \1\,,  \label{sol-fuzz}\\
&& X^5 = \1\otimes \1\otimes \left(\frac{4}{3}\mu\cdot c\sqrt{\frac{\al}{2}} 
J^1\right)\otimes \1\,,
 \quad 
X^6 = \1\otimes \1\otimes\left(\frac{4}{3}\mu\cdot c\sqrt{\frac{\al}{2}}
J^2\right)\otimes \1\,, 
\nn 
\\ 
&& X^7 = \1\otimes \1\otimes \1\otimes \left(\frac{4}{3}\mu\cdot 
c\sqrt{\frac{1-\al}{2}}
J^1\right)\,, \quad 
X^8 = \1\otimes \1\otimes \1 \otimes \left(\frac{4}{3}\mu\cdot c\sqrt{\frac{1-\al}{2}}
J^2\right)\,, \nn \\
&& \omega = \left(2\mu\cdot s J^1\right) \otimes 
\left(-\mu J^1\right) \otimes \1 \otimes \1\,, \quad 
 D_{\tau}X^5 = \cdots D_{\tau}X^9 = \psi =0\,. \nn
\end{eqnarray}
The symbol $\otimes$ means the embedding of the $SU(2)$ generators into
the matrix, and the $i$th generator belongs to the $N_i$-dimensional
representation where $N = N_1 + N_2 + N_3 + N_4$.  For example,
$\left(\frac{4}{3}\mu J^3\right) \otimes \1 \otimes
\left(c\cdot\frac{4}{3}\mu J^3\right)\otimes \left(c\cdot \frac{4}{3}\mu
J^3\right)$ is represented by the following matrix 
\begin{eqnarray}
\left(\frac{4}{3}\mu J^3\right)
\otimes \1 \otimes \left(c\cdot\frac{4}{3}\mu J^3\right)\otimes 
\left(c\cdot \frac{4}{3}\mu J^3\right) 
=
\left(
  \begin{array}{cccc}
\frac{4}{3}\mu J^3 &  &  &\\
 & \1 & & \\
 &  & c\cdot\frac{4}{3}\mu J^3  &  \\
 & &   & c\cdot\frac{4}{3}\mu J^3 
  \end{array}
\right)
%
\begin{array}{c}
\Big\} N_1 \\
\Big\} N_2 \\
\Big\} N_3 \\
\Big\} N_4
\end{array}
\,. \nn 
\end{eqnarray}
The above solution (\ref{sol-fuzz}) preserves 4 dynamical
supersymmetries while the 16 kinematical supersymmetries are
broken.  Hence this is a 1/5(=4/20) BPS object\footnote{In the pp-wave
matrix model, such fuzzy sphere solutions preserves 16 dynamical
supersymmetries and are the 1/2(=16/32) BPS objects. }.  In our case
this configuration has zero energy as in the pp-wave matrix model (i.e.,
$V=0$).  Therefore, this configuration can appear in the classical
vacuum without the cost of energy, although this is not a 
classical solution of equation of motion.

\section{Conclusion and Discussion}

We have discussed a matrix model on a homogeneous plane-wave
background. The action of the matrix model has been proposed. This
theory has 20 supersymmetries, and we have explicitly constructed the
transformation laws of 4 dynamical and 16 kinematical supersymmetries in
this model. Then, the vacuum energy of the abelian part has been
calculated, and we have shown that the vacuum energy in the
$P\neq 0$ is identical with that in the $P=0$ case. In particular, the
effect of a parameter $P$ is to shift the origin of the harmonic oscillator
$X^2$ and to rescale the continuous spectrum of $X^1$. We have also found
classical solutions. The graviton solution is slightly different from
that in the usual pp-wave cases because of the presence of the
additional parameter $P$. It has been shown that the fuzzy hyperboloid
and fuzzy sphere solutions are also exist in our model. Notably, 
the fuzzy ellipsoidal sphere solution (\ref{FS}) with (\ref{fz1})
and the  fuzzy hyperboloid solution (\ref{FS3}) with (\ref{fz4}) 
have negative energies.

It would also be interesting to study a supermembrane theory by taking a
large $N$ limit (for the pp-wave case, see \cite{SY1}).  By assuming the
matrix regularization \cite{dWHN}, the action of supermembrane theory can formally
be written down as follows:
\begin{eqnarray}
S &=& \int\!d\tau d^2\sigma w(\sigma)\,\frac{1}{2}\Biggl[(D_{\tau}X^i)^2 -
 \frac{1}{2}\{X^i,X^j\}^2 
+ i\psi^TD_{\tau}\psi -
 i\psi^{T}\gamma_i\{X^i,\psi\}  \nn \\
&& \qquad -4\mu^2(1-P^2)(X^2)^2 - \mu^2(X^3)^2 -\mu^2(X^4)^2 
+ 2\mu P X^1D_{\tau}X^2 - 2\mu P X^2D_{\tau}X^1 \nn \\
&& \qquad + \frac{\mu}{2}i\psi^T \left(
W - P\gamma^{12}
\right) \psi + \frac{2}{3}\mu \cdot \Bigl\{\sum_{I,J,K=1,2,9} 
- P \sum_{I,J.K=3,4,9} \nn \\
&& \qquad + (1-P^2)^{1/2} \sum_{I,J,K=2,5,6} 
+ (1-P^2)^{1/2} \sum_{I,J,K=2,7,8}
\Bigr\}\epsilon_{IJK}X^{K}\{X^I,X^J\}
\Biggr]\,, \\
W &=& \gamma^{129} - P \gamma^{349} + (1-P^2)^{1/2}\gamma^{256} + 
(1-P^2)^{1/2}\gamma^{278}\,,\nn
\end{eqnarray}
where the covariant derivative is $D_{\tau}\ast = \partial_{\tau}\ast +
\{\omega,\ast\}$ and $\{A,\,B\}\equiv (\epsilon_{ab}/w)\partial_a A
\partial_b B$ is the Lie bracket.  The indices $a,b$ represent the
spacial directions of membrane world-volume. An interesting subject to
study is the brane charge in our model. The brane charges are modified
in the nontrivial background \cite{SY1,HS1}.  In addition, it would be
interesting to study an open supermembrane theory on the time-dependent
plane-wave background as in \cite{SY1,SaYo1} because this background has
quite special properties
and so the classification of Dirichlet branes are
also nontrivial. All of the above topics will be reported in our next
paper \cite{SSY}.

One of the most interesting subjects is the stability of a single
supermembrane on the time-dependent background.  In the flat case a
single supermembrane is unstable as is well known \cite{dWLN}. This fact
is deeply based on the cancellation of zero-point energies between
bosons and fermions. However, it is not clear in general whether the
bosonic zero-point energy cancels out the fermionic one. 
Remarkably, our result on the vacuum of the abelian part of the matrix
model here should support that the zero-point energies cannot cancel
out. Moreover, the flat directions are left in the time-dependent
homogeneous plane-wave in contrast with the maximally supersymmetric
pp-wave case.  This issue is an interesting future work.

Finally, we comment on the relation of our matrix model to the
time-dependent background as a closing remark. The homogeneous
plane-wave background we have considered can be mapped to the explicitly
time-dependent plane-wave background via the time-dependent coordinate
transformation. However, once we move to the stationary frame
in which 
the
time dependence is not manifest, the effect of the time dependence can
be expressed by a kind of the vector potential, which leads to the
constant magnetic field. Thus, we might expect the relationship of our
model to the noncommutative geometry or Seiberg-Witten map \cite{SW}. In
addition, the plane-wave geometry considered here is closely related to
the Lewis-Riesenfeld invariant theory \cite{LR} as noted in
\cite{BO}. This theory leads us to the quantization of the system of
time-dependent harmonic oscillators. Study of such a system would be a
useful laboratory to promote the understanding of the time-dependent
backgrounds and to develop the techniques to treat theories on them.  
We believe that our model should be a clue to shed light on the physics 
on the time-dependent backgrounds.

\acknowledgments

The authors would like to thank Katsuyuki Sugiyama for 
useful discussion at the first stage of this work.


\appendix 


\section{Matrix model on the general homogeneous plane-wave}
\label{general:app}

Here we shall discuss the matrix model on the general homogeneous
plane-wave background (anti-Mach space) whose metric is (\ref{2.1}) and
the four-form flux (\ref{2.2}) is equipped. As discussed in \cite{BMO},
this background admits extra Killing spinors satisfying
\begin{eqnarray}
&& [
\gamma_i\theta^2
+6\theta\gamma_i\theta
+9\theta^2\gamma_i
+6\gamma_i[\theta,\phi]
+18[\theta,\phi]\gamma_i 
-144(A_{ij}\gamma^j+f_{ij}f_{jk}\gamma^k)
]\epsilon=0\,,
\label{projection.bg}
\\
&&\epsilon(x^+)=
e^{\frac{x^+}{2}(\phi+\frac{1}{6}W)}\epsilon_0
\label{t-derivative.bg}
\end{eqnarray}
where we have defined as 
\begin{eqnarray}
\theta \equiv \frac{1}{3!}\xi_{ijk}\gamma^{ijk}\,,\quad 
\phi \equiv \frac{1}{2}f_{ij}\gamma^{ij}\,,
\end{eqnarray}
and $\epsilon_0$ is a constant spinor.

The matrix theory action on the  background is given by 
\begin{eqnarray}
S&=&\frac{1}{2}\int\!\! d\tau\,{\rm Tr}\Big[
(D_\tau X^i)^2
+\frac{1}{2}[X^i,X^j]^2
+\mu^2A_{ij}X^iX^j
-2\mu f_{ij}X^iD_\tau X^j
\nonumber\\&&~~~~~~~~
+i\psi^T D_\tau\psi
-\psi^T\gamma_i[X^i,\psi]
+\frac{\mu}{2}i\psi^T(W+\phi)\psi
+\frac{2}{3}i\mu\xi_{ijk}X^iX^jX^k
\Big]\,,
\label{action;general}
\end{eqnarray}
where the matrix $W$ and the covariant derivative are defined as,
respectively,  
\begin{eqnarray}
W \equiv -\frac{1}{2}\,\theta\,, \quad D_\tau *=\partial_\tau *-i[\omega, *]\,.
\end{eqnarray}
The action (\ref{action;general})
is invariant under the dynamical supersymmetry transformations
\begin{eqnarray}
\delta_\epsilon X^i&=&
 i\psi^T\gamma^i\epsilon\,,\quad \delta_\epsilon w\;=\;
 i\psi^T\epsilon\,,\\
\delta_\epsilon\psi&=&
 \Big[
  D_\tau X^i\gamma_i-2\mu
  f_{ij}X^i\gamma^j-\frac{i}{2}[X^i,X^j]\gamma_{ij} \nn \\
&& \quad  -\frac{\mu}{6}X^i\gamma_i(W-3\phi)
 -\frac{\mu}{2}X^i(W+\phi)\gamma_i\Big]\epsilon\,,
\end{eqnarray}
provided that
\begin{eqnarray}
&&
[
\gamma_iW^2
+6W\gamma_iW
+9W^2\gamma_i
-3\gamma_i[W,\phi]
-9[W,\phi]\gamma_i \nn \\
&& \qquad \qquad -36(A_{ij}\gamma^j+f_{ij}f_{jk}\gamma^k)
]\epsilon=0,
\label{projection}
\\&&
\epsilon= e^{\frac{\mu}{6}(W-3\phi)\tau}\epsilon_0.
\label{t-derivative}
\end{eqnarray}
The Eqs.\,(\ref{projection}) and (\ref{t-derivative}) are nothing but
(\ref{projection.bg}) and (\ref{t-derivative.bg}) respectively. The
action (\ref{action;general}) is also invariant under the kinematical
supersymmetries
\begin{eqnarray}
\delta_\eta X^i=\delta_\eta\omega =0,~~~\delta_\eta\psi=\eta (\tau ),~~~
\eta(\tau )=e^{-\frac{\mu}{2}(W+\phi)\tau}\eta_0
\end{eqnarray}
where $\eta_0$ is a constant spinor.

\section{Analysis of $X^1$ and $X^2$ directions in the $P\neq 0$ case} 
\label{derive:app}

Here we will examine the $X^1$ and $X^2$ directions in the $P\neq 0$
case where the analysis of these directions are quite nontrivial and 
complicated due to the presence of the nonzero $f_{ij}$. 
The Lagrangian of this part is given by 
\begin{eqnarray}
L = \frac{1}{2}\left[
(\dot{X}^1)^2 + (\dot{X}^2)^2 - 4\mu^2(1-P^2)(X^2)^2 +2\mu P 
(X^1\dot{X}^2 - X^2\dot{X}^1) 
\right]\,.
\end{eqnarray}
By using the standard procedure, the Hamiltonian can be 
derived as follows:
\begin{eqnarray}
H = \frac{1}{2}(\pi^1 + \mu P X^2)^2 + \frac{1}{2}(\pi^2 - \mu P X^1)^2 
+ 2\mu^2 (1-P^2)(X^2)^2\,.
\end{eqnarray}
Here we quantize the system by imposing the commutation relation:
\[
 [X^i,\pi^j] = i\del^{ij}\,,
\]
and will solve the energy eigenvalue problem:
\[
 H \Psi = E\Psi\,.
\]
When we decompose the $\Psi$ as $\Psi = \e^{+i\mu P X^1X^2}\chi$, 
we can obtain the equation for the $\chi$ described as 
\begin{eqnarray}
\tilde{H}\chi &=& E\chi \,, \nn \\
\tilde{H} &\equiv& \frac{1}{2}\left(\pi^1 + 2\mu P X^2\right)^2 
+ \frac{1}{2}(\pi^2)^2 
+ 2\mu^2 (1-P^2)(X^2)^2\,.  
\end{eqnarray}
The expression $\chi = \e^{ip_1X^1}\phi$, where $p_1$ is
the momentum for the $X^1$-direction, allows us to rewrite 
the Hamiltonian $\tilde{H}$ as 
\begin{eqnarray}
\tilde{H} = \frac{1}{2}(\pi^2)^2 + \frac{1}{2}(2\mu)^2
\left(X^2 + \frac{1}{2\mu}p_1P
\right)^2 + \frac{1}{2}(p_1)^2(1-P^2)\,.
\end{eqnarray}
Thus, we can see that the effect of a parameter $P$ is 
to shift the origin of harmonic oscillation of $X^2$-direction and 
to rescale the kinetic energy (continuous spectrum) of $X^1$. 

Finally. we would like to note that the final expression of Hamiltonian
is closely related to the invariant in the context of the Lewis and
Riesenfeld theory \cite{LR}, which can quantize the system of time-dependent
harmonic oscillator.

\end{document}